\documentstyle [12pt] {article}

\catcode`@=12
\begin{document}
\thispagestyle{empty}
\begin{flushright} UCRHEP-T274\\April 2000\
\end{flushright}
\vskip 0.5in
\begin{center}
{\large \bf Three Inequivalent Mass-Degenerate Majorana Neutrinos\\
and a Model of Their Splitting for Neutrino Oscillations\\}
\vskip 2.0in
{\bf Rathin Adhikari$^1$, Ernest Ma$^2$, G. Rajasekaran$^1$\\}
\vskip 0.3in
{$^1$ \sl Institute of Mathematical Sciences, Chennai (Madras) 600113, India\\}
\vskip 0.1in
{$^2$ \sl Physics Department, Univeristy of California, Riverside, 
CA 92521, USA\\}
\end{center}
\vskip 1.2in
\begin{abstract}\
The mass matrix of three Majorana neutrinos of equal mass is not necessarily 
proportional to the identity matrix, but expressible in terms of two angles 
and one phase.  We discuss how such a mass matrix may be stable or unstable 
against radiative corrections.  We then propose a model with additional 
explicit breaking of the threefold degeneracy to account for the atmospheric 
neutrino data, while the radiative breaking explains the solar neutrino data, 
using the large-angle Mikheyev-Smirnov-Wolfenstein solution.  Our model 
requires a nonzero effective $\nu_e$ mass for neutrinoless double beta decay 
close to the present experimental upper limit of 0.2 eV.
\end{abstract}

\newpage
\baselineskip 24pt

There are now several experimental results\cite{atm,sol,lab} favoring 
neutrino oscillations as their explanation.  Since only the differences of 
the squares of neutrino masses are involved, there has always been a lot 
of theoretical interest in considering nearly degenerate neutrino 
masses\cite{degenerate}.  The origin of their splitting may in fact be 
radiative\cite{radiative} and some simple specific models have been 
proposed\cite{permute}.  In the case of 3 Majorana neutrinos, it has been 
shown\cite{three} that all may be identical in mass and yet are inequivalent 
to one another because their mass matrix may contain 2 angles and 1 phase 
which cannot be redefined away.  In the following we will discuss the 
stability of this construction under radiative corrections.

We will then propose a model with additional explicit breaking of the 
threefold degeneracy to account for the atmospheric neutrino data\cite{atm}, 
while the radiative breaking explains the solar neutrino data\cite{sol}, 
using the large-angle Mikheyev-Smirnov-Wolfenstein (MSW) solution\cite{msw}.  
Our model assumes all 3 neutrino masses to be of order 0.5 eV, to be 
consistent\cite{structure} with their possible role in cosmic structure 
formation, in the light of recent astrophysical evidence\cite{cosmo} 
favoring a nonzero cosmological constant. An effective $\nu_e$ mass for 
neutrinoless double beta decay close to the present experimental upper 
limit\cite{double} of 0.2 eV is also required.

We start out with the most general $3 \times 3$ unitary matrix linking 
$(\nu_e, \nu_\mu, \nu_\tau)$ with their mass eigenstates $(\nu_1, \nu_2, 
\nu_3)$, i.e.\cite{adhraj}
\begin{equation}
U = \left[ \begin{array} {c@{\quad}c@{\quad}c} c_1 c_3 & 
s_1 c_3 e^{-i \delta_1} & s_3 e^{-i\delta_2} \\ -s_1 c_2 
e^{i\delta_1} - c_1 s_2 s_3 e^{i(\delta_2 + \delta_3)} & c_1 c_2  - 
s_1 s_2 s_3 e^{i(\delta_3 + \delta_2 - \delta_1)} & s_2 c_3 
e^{i\delta_3} \\ s_1 s_2 e^{i(\delta_1 - \delta_3)} 
- c_1 c_2 s_3 e^{i\delta_2} & -c_1 s_2 e^{-i\delta_3} - 
s_1 c_2 s_3 e^{i(\delta_2 - \delta_1)} & c_2 c_3 \end{array} \right],
\end{equation}
which has 3 angles and 3 phases.  Let us set $s_3 = 0$ ($c_3 = 1$), 
$\delta_1 = \pi/2$, $\delta_2 \equiv \delta$, and $\delta_3 = \pi/2 - 
\delta$, and multiply on the left by the diagonal matrix $(1, -i, 
e^{-i\delta})$, then
\begin{equation}
U = \left[ \begin{array} {c@{\quad}c@{\quad}c} c_1 & -i s_1  & 0 \\ 
- s_1 c_2 & -i c_1 c_2 & s_2 e^{-i\delta} \\ s_1 s_2 & i c_1 s_2  & c_2 
e^{-i\delta} \end{array} \right].
\end{equation}

Since all neutrino masses are assumed equal, the mass matrix in the 
$(\nu_1, \nu_2, \nu_3)$ basis is just $m$ times the identity matrix. 
However, in the $(\nu_e, \nu_\mu, \nu_\tau)$ basis, it is given by
\begin{equation}
{\cal M} = m U^* U^\dagger = m \left[ \begin{array} {c@{\quad}c@{\quad}c} 
c_0 & - s_0 c_2 & s_0 s_2 \\ - s_0 c_2 & - c_0 c_2^2 + s_2^2 e^{2i\delta} & 
s_2 c_2 (c_0 + e^{2i\delta}) \\ s_0 s_2 & s_2 c_2 (c_0 + e^{2i\delta}) & 
-c_0 s_2^2 + c_2^2 e^{2i\delta} \end{array} \right],
\end{equation}
where $c_0 \equiv c_1^2 - s_1^2$ and $s_0 \equiv 2 s_1 c_1$.  As shown in 
Ref.\cite{three}, this mass matrix (which has 2 angles and 1 phase) cannot 
be reduced further and there is $CP$ violation if $e^{2i\delta} \neq \pm 1$.  
Another way to see this is to realize that $U$ of Eq.~(2) cannot be written 
in the form $U = D O$, where $D$ is a diagonal matrix containing phases and 
$O$ is an orthogonal matrix.

Even though $U$ of Eq.~(2) is nontrivial, the absence of any mass differences 
among the 3 neutrinos will not result in oscillations.  However, in the 
presence of radiative corrections from the charged leptons, neutrino mass 
differences will occur.  It is easiest to do this in the $(\nu_1, \nu_2, 
\nu_3)$ basis.  Without radiative corrections, this mass matrix is just 
$m$ times the identity matrix.  With radiative corrections coming from the 
wavefunction renormalization of $\nu_\tau$ due to $m_\tau$, we find
\begin{eqnarray}
{\cal M}' &=& m U^T \left[ \begin{array} {c@{\quad}c@{\quad}c} 1 & 0 & 0 \\ 
0 & 1 & 0 \\ 0 & 0 & 1 + \epsilon \end{array} \right] U^* U^\dagger \left[ 
\begin{array} {c@{\quad}c@{\quad}c} 1 & 0 & 0 \\ 0 & 1 & 0 \\ 0 & 0 & 1 + 
\epsilon \end{array} \right] U \nonumber \\ &=& m \left[ \begin{array} 
{c@{\quad}c@{\quad}c} 1 + 2 s_1^2 s_2^2 \epsilon & 0 & 2 c_\delta s_1 s_2 
c_2 \epsilon \\ 0 & 1 + 2 c_1^2 s_2^2 \epsilon & -2 s_\delta c_1 s_2 c_2 
\epsilon \\ 2 c_\delta s_1 s_2 c_2 \epsilon & -2 s_\delta c_1 s_2 c_2 
\epsilon & 1 + 2 c_2^2 \epsilon \end{array} \right],
\end{eqnarray}
where
\begin{equation}
\epsilon = - {G_F m_\tau^2 \over 16 \pi^2 \sqrt 2} \ln {\Lambda^2 \over 
m_W^2},
\end{equation}
with $\Lambda$ equal to the scale at which the degenerate mass matrix is 
defined.

The characteristic polynomial equation of ${\cal M}'/m$ is given by
\begin{equation}
x^3 - x^2 + x [ s_2^2 c_2^2 (1 - c_\delta^2 
s_1^2 - s_\delta^2 c_1^2) + s_1^2 c_1^2 s_2^4] = 0,
\end{equation}
where $x = (\lambda - 1)/2\epsilon$. Hence one eigenvalue $\lambda$ is 1 and 
it corresponds to the eigenstate
\begin{equation}
\nu' = N^{-1/2} [c_\delta c_1 c_2 \nu_1 - s_\delta s_1 c_2 \nu_2 - 
s_1 c_1 s_2 \nu_3] = N^{-1/2} [c_2 (c_\delta c_1^2 - i s_\delta s_1^2) \nu_e 
- s_1 c_1 e^{i\delta} \nu_\mu],
\end{equation}
where $N = c_\delta^2 c_1^2 c_2^2 + s_\delta^2 s_1^2 c_2^2 + 
s_1^2 c_1^2 s_2^2$.  This shows that the original pattern of 2 mixing angles 
and 1 phase is generally unstable and no matter what values they take, 
a mass eigenstate exists without $\nu_\tau$ as given above.  For ${\cal M}'$ 
to be stable, the off-diagonal elements must go to zero, but since they 
depend on the original 2 angles and 1 phase, the condition of stability 
imposes severe constraints on their values.

To understand the atmospheric neutrino data in terms of oscillations, we 
need $s_2^2 \sim c_2^2 \sim 0.5$, hence the requirement of stability forces 
$s_1$ (as well as $s_\delta$) to be small, which implies $\nu' \simeq \nu_1 
\simeq \nu_e$.  This means that we must take the small-angle MSW solution 
for the solar neutrino data.  However from Eq.~(6), it can be shown that 
the mass eigenvalues of the other 2 states are both smaller than 1 because 
$\epsilon$ of Eq.~(5) is negative.  Thus there cannot be any resonance 
enhancement of oscillations from $\nu_e$ interactions in the sun and this 
scheme does not work.

We now propose to break the threefold degeneracy also explicitly by a mass 
term
\begin{equation}
m' (s' \nu_2 + c' \nu_3)^2,
\end{equation}
where $m'$ is chosen such that $(m+m')^2 - m^2$ is of order $10^{-3}$ eV$^2$ 
to be suitable for atmospheric neutrino oscillations.  Then the $2 \times 2$ 
mass submatrix spanning $\nu_1$ and $c' \nu_2 - s' \nu_3$ is given by
\begin{equation}
{\cal M}'' = m \left[ \begin{array} {c@{\quad}c} 1 + 2 s_1^2 s_2^2 \epsilon 
& -2 s' c_\delta s_1 s_2 c_2 \epsilon \\ -2 s' c_\delta s_1 s_2 c_2 \epsilon 
& 1 + 2 (c'^2 c_1^2 s_2^2 + s'^2 c_2^2 + 2 s_\delta s' c' c_1 s_2 c_2) 
\epsilon \end{array} \right].
\end{equation}
Consider the simplified case of $s_\delta = -1$ ($c_\delta = 0$), then 
${\cal M}''$ is diagonal.  Hence $\nu_1$ and $c' \nu_2 - s' \nu_3$ are 
eigenstates with eigenvalues
\begin{equation}
m_1 = m (1 + 2 s_1^2 s_2^2 \epsilon), ~~~ m_2 = m [1 + 2 (c'^2 s_2^2 c_1^2 + 
s'^2 c_2^2 - 2 s' c' s_2 c_2 c_1) \epsilon]
\end{equation}
respectively.  We choose the convention that $c_1^2 > 1/2$, so that $\nu_1$ 
is mostly $\nu_e$.  Therefore, the requirement $m_1 < m_2$ means
\begin{equation}
s_1^2 s_2^2 > c'^2 s_2^2 c_1^2 + s'^2 c_2^2 - 2 s' c' s_2 c_2 c_1 = 
(c' s_2 c_1 - s' c_2)^2,
\end{equation}
resulting in the following condition on $c_1$:
\begin{equation}
{1 \over \sqrt 2} < c_1 < {s' c' c_2 + \sqrt {2 s_2^2 - s'^2} \over s_2 
(1 + c'^2)}.
\end{equation}
Note that $s' = 0$ is not a solution.

With the new set of mass eigenstates, the transformation matrix $U$ of Eq.~(2) 
becomes
\begin{equation}
U_{\alpha i} = \left[ \begin{array} {c@{\quad}c@{\quad}c} c_1 & -i c' s_1 & 
-i s' s_1 \\ -s_1 c_2 & -i (c' c_1 c_2 + s' s_2) & -i (s' c_1 c_2 - c' s_2) 
\\ s_1 s_2 & i (c' c_1 s_2 - s' c_2) & i (s' c_1 s_2 + c' c_2) \end{array} 
\right].
\end{equation}
Using the well-known expressions for the probabilities of neutrino 
oscillations, we find in the atmospheric case,
\begin{eqnarray}
P_{\mu \mu} &=& 1 - 2 |U_{\mu 3}|^2 (1 - |U_{\mu 3}|^2) (1- \cos (\Delta m^2 L 
/2E)) \\ P_{ee} &=& 1 - 2 |U_{e3}|^2 (1 - |U_{e3}|^2) (1- \cos (\Delta m^2 L 
/2E)) \\ P_{\mu e} &=& P_{e \mu} ~=~ 2 |U_{e3}|^2 |U_{\mu 3}|^2 (1 - \cos 
(\Delta m^2 L/2E)),
\end{eqnarray}
where $\Delta m^2 = (m+m')^2 - m^2$.  In the solar case,
\begin{equation}
P_{ee} = 1 - 2 |U_{e3}|^2 (1 - |U_{e3}|^2) - 2 |U_{e1}|^2 |U_{e2}|^2 
(1 - \cos (\Delta' m^2 L /2E)),
\end{equation}
where $\Delta' m^2 = m_2^2 - m_1^2$.  Consider again the small-angle MSW 
solution of the solar neutrino data.  This requires $c_1 \simeq 1$, but then 
Eq.~(11) implies that $(c' s_2 - s' c_2)^2 \simeq |U_{\mu 3}|^2$ has to be 
very small, which is disallowed by the atmospheric neutrino data.  Note 
especially that this restriction is independent of the value of $m$.  Hence 
only the large-angle MSW solution\cite{large} will be considered from now on.

We allow $\epsilon$ of Eq.~(5) to be divided by $\cos^2 \beta$, where 
$\tan \beta \equiv v_2/v_1$ in a two-Higgs doublet model.  We then fix
\begin{eqnarray}
\Delta' m^2 = m_2^2 - m_1^2 \simeq {4 m^2 G_F m_\tau^2 \over 16 \pi^2 \sqrt 2 
\cos^2 \beta} \ln {\Lambda^2 \over m_W^2} \left[ s_1^2 s_2^2 - (c' s_2 c_1 - 
s' c_2)^2 \right]
\end{eqnarray}
to be equal to $10^{-5}$ eV$^2$ to fit the solar data.  This is necessary 
because we will set $\Lambda = 10^{14}$ GeV and choose $m$ to be 0.5 eV or 
less\cite{structure} which then require $\cos^2 \beta < 1$ to obtain the 
desired value of $\Delta' m^2$.  From the CHOOZ reactor data\cite{chooz}, 
we require
\begin{equation}
|U_{e3}|^2 < 0.025.
\end{equation}
From the atmospheric neutrino data\cite{atm}, we require
\begin{equation}
0.84 < 4 |U_{\mu 3}|^2 (1 - |U_{\mu 3}|^2) = \sin^2 2 \theta_{atm} < 1.
\end{equation}
From the large-angle MSW solution\cite{large} to the solar neutrino 
data\cite{sol}, we require
\begin{equation}
0.7 < 4 |U_{e1}|^2 |U_{e2}|^2 = \sin^2 2 \theta_{sol} < 0.9.
\end{equation}
From neutrinoless double beta decay\cite{double}, we require
\begin{equation}
m_{ee} = m c_0 = m (c_1^2 - s_1^2) < 0.2 ~{\rm eV}.
\end{equation}
We then choose $m$ = 0.5 eV, and scan the parameter space 
of $s'$, $s_1$, and $s_2$ for solutions.  We find the following allowed 
ranges of values (each obtained with the others free):
\begin{equation}
0.16 < s' < 0.29, ~~~ 0.55 < s_1 < 0.64, ~~~ 0.66 < s_2 < 0.84.
\end{equation}
In Table 1 we show some typical solutions.  It is clear that the effective 
neutrino mass $m_{ee}$ for neutrinoless double beta decay is not far below 
its current experimental upper limit of 0.2 eV and is accessible to the 
next generation of such experiments\cite{next}.  We note also that Eq.~(18) 
depends only on the ratio $m/\cos \beta$.  Hence if we decrease $m$, then 
the same set of values for $s'$, $s_1$, and $s_2$ is a solution if we also 
decrease $\cos \beta$ by the same factor.  Of course, $m_{ee}$ is also 
reduced, but if $m$ plays a role in cosmic structure formation, then $m_{ee}$ 
cannot be an order of magnitude smaller than 0.1 eV.

In the above, we have chosen $s_\delta = -1$ in Eq.~(9).  This allows 
Eq.~(11) to be satisfied with the widest range of parameter values.  
For a general $s_\delta$, Eq.~(11) is replaced by
\begin{equation}
s_1^2 s_2^2 > c'^2 s_2^2 c_1^2 + s'^2 c_2^2 + 2 s_\delta s' c' s_2 c_2 c_1 
= (c' s_2 c_1 - s' c_2)^2 + 2 (1+s_\delta) s' c' s_2 c_2 c_1,
\end{equation}
which is clearly more restrictive.

In conclusion, we have shown in this paper how three inequivalent 
mass-degenerate Majorana neutrinos are unstable against radiative corrections 
due to $m_\tau$.  However, if we add an explicit mass term which breaks 
this threefold degeneracy to account for the atmospheric neutrino data, the 
remaining radiative splitting is able to account for the solar data, but 
only with the large-angle MSW solution, resulting in an effective neutrino 
mass for neutrinoless double beta decay close to the present experimental 
upper limit of 0.2 eV.

The research of E.M. was supported in part by the U.~S.~Department of Energy 
under Grant No.~DE-FG03-94ER40837.

\newpage
\bibliographystyle{unsrt}

\begin{thebibliography}{99}

\bibitem{atm} Y. Fukuda {\it et al.}, Phys. Lett. {\bf B433}, 9 (1998); 
{\bf B436}, 33 (1998); Phys. Rev. Lett. {\bf 81}, 1562 (1998); {\bf 82}, 
2644 (1999).
\bibitem{sol} R. Davis, Prog. Part. Nucl. Phys. {\bf 32}, 13 (1994); 
P. Anselmann {\it et al.}, Phys. Lett. {\bf B357}, 237 (1995); {\bf B361}, 
235 (1996); J. N. Abdurashitov {\it et al.}, Phys. Lett. {\bf B328}, 234 
(1994); Y. Fukuda {\it et al.}, Phys. Rev. Lett. {\bf 77}, 1683 (1996); 
{\bf 81}, 1158 (1998); {\bf 82}, 1810 (1999); {\bf 82}, 2430 (1999).
\bibitem{lab} C. Athanassopoulos {\it et al.}, Phys. Rev. Lett. {\bf 75}, 
2650 (1995); {\bf 77}, 3082 (1996); {\bf 81}, 1774 (1998).
\bibitem{degenerate} D. Caldwell and R. N. Mohapatra, Phys. Rev. {\bf D48}, 
3259 (1993);  
A. S. Joshipura, Z. Phys. {\bf C64}, 31 (1994); Phys. Rev. {\bf D51}, 
1321 (1995); 
P. Bamert and C. P. Burgess, Phys. Lett. {\bf B329}, 289 (1994); 
D.-G. Lee and R. N. Mohapatra, Phys. Lett. {\bf B329}, 463 (1994); 
A. Ioannisian and J. W. F. Valle, Phys. Lett. {\bf B332}, 93 (1994); 
A. Ghosal, Phys. Lett. {\bf B398}, 315 (1997); 
A. K. Ray and S. Sarkar, Phys. Rev. {\bf D58}, 055010 (1998); 
C. D. Carone and M. Sher, Phys. Lett. {\bf B420}, 83 (1998); 
F. Vissani, hep-ph/9708483; 
H. Georgi and S. L. Glashow, Phys. Rev. {\bf D61}, 097301 (2000); 
U. Sarkar, Phys. Rev. {\bf D59}, 037302 (1999); 
R. N. Mohapatra and S. Nussinov, Phys. Rev. {\bf D60}, 013002 (1999); 
G. C. Branco, M. N. Rebelo, and J. I. Silva-Marcos, hep-ph/9906368;
C. Wetterich, Phys. Lett. {\bf B451}, 397 (1999); 
Y. L. Wu, Phys. Rev. {\bf D60}, 073010 (1999); Eur. 
Phys. J. {\bf C10}, 491 (1999); Int. J. Mod. Phys. {\bf A14}, 4313 (1999); 
hep-ph/9906435; 
R. Barbieri, L. J. Hall, G. L. Kane, and G. G. Ross, hep-ph/9901228;
M. Tanimoto, T. Watari, and T. Yanagida, Phys. Lett. {\bf B461}, 345 (1999);
A. K. Ray and S. Sarkar, Phys. Rev. {\bf D61}, 035007 (2000).
\bibitem{radiative} E. Ma, Phys. Lett. {\bf B456}, 48 (1999); {\bf B456}, 
201 (1999); 
J. Phys. {\bf G25}, L97 (1999); hep-ph/9907503; 
J. Ellis and S. Lola, Phys. Lett. {\bf B458}, 310 (1999); 
J. A. Casas, J. R. Espinosa, A. Ibarra, and I. Navarro, Nucl. Phys. 
{\bf B556}, 3 (1999); J. High Energy Phys. {\bf 9}, 015 (1999); 
Nucl. Phys. {\bf B569}, 82 (2000); hep-ph/9910420; 
R. Barbieri, G. G. Ross, and A. Strumia, J. High Energy Phys. {\bf 10}, 020 
(1999);
Y. L. Wu, hep-ph/9905222; 
N. Haba and N. Okamura, hep-ph/9906481;
N. Haba, Y. Matsui, N. Okamura, and M. Sugiura, hep-ph/9908429; 
hep-ph/9911481;
P. H. Chankowski, W. Krolikowski, and S. Pokorski, Phys. Lett. {\bf B473}, 
109 (2000);
K. R. S. Balaji, A. S. Dighe, R. N. Mohapatra, and M. K. Parida, 
hep-ph/0001310; hep-ph/0002177.

\bibitem{permute} E. Ma, Phys. Rev. Lett. {\bf 83}, 2514 (1999); Phys. Rev. 
{\bf D61}, 033012 (2000); 
R. N. Mohapatra, A. Perez-Lorenzana, and C. A. de S. Pires, Phys. Lett. 
{\bf B474}, 355 (2000); 
E. J. Chun and S. K. Kang, hep-ph/9912524.

\bibitem{three} G. C. Branco, M. N. Rebelo, and J. I. Silva-Marcos, Phys. Rev. 
Lett. {\bf 82}, 683 (1999).

\bibitem{msw} L. Wolfenstein, Phys. Rev. {\bf D17}, 2369 (1978); S. P. 
Mikheyev and A. Yu. Smirnov, Sov. J. Nucl. Phys. {\bf 42}, 913 (1986).

\bibitem{structure} M. Fukugita, G.-C. Liu, and N. Sugiyama, Phys. Rev. Lett. 
{\bf 84}, 1082 (2000).

\bibitem{cosmo} A. G. Riess {\it et al.}, Astron. J. {\bf 116}, 1009 (1998); 
S. Perlmutter {\it et al.}, astro-ph/9812133.

\bibitem{double} L. Baudis {\it et al.}, Phys. Rev. Lett. {\bf 83}, 41 (1999).

\bibitem{adhraj} R. Adhikari and G. Rajasekaran, Phys. Rev. {\bf D61}, 
031301 (2000).

\bibitem{large} J. N. Bahcall, P. I. Krastev, and A. Yu. Smirnov, Phys. Rev. 
{\bf D60}, 093001 (1999).

\bibitem{chooz} M. Apollonia {\it et al.}, Phys. Lett. {\bf B466}, 415 (1999).

\bibitem{next} H. V. Klapdor-Kleingrothaus, hep-ex/9907040.

\end{thebibliography}

\newpage
\begin{table}
\begin{center}
\begin{tabular}{|c|c|c|c|c|c|c|}
\hline
$s'$ & $s_1$ & $s_2$ & $\cos \beta$ & $\sin^2 2 \theta_{atm}$ & 
$\sin^2 2 \theta_{sol}$ & $m_{ee}$ \\
\hline
0.18 & 0.60 & 0.68 & 0.06 & 0.87 & 0.89 & 0.14 eV\\
\hline
0.23 & 0.58 & 0.72 & 0.07 & 0.88 & 0.85 & 0.16 eV\\
\hline
0.23 & 0.62 & 0.72 & 0.21 & 0.89 & 0.90 & 0.12 eV\\
\hline
0.23 & 0.62 & 0.76 & 0.18 & 0.95 & 0.90 & 0.12 eV\\
\hline
0.23 & 0.62 & 0.80 & 0.14 & 0.99 & 0.90 & 0.12 eV\\
\hline
0.28 & 0.56 & 0.74 & 0.10 & 0.85 & 0.79 & 0.19 eV\\
\hline
\end{tabular}
\caption{Typical allowed parameter values of this model for $m$ = 0.5 eV.}
\end{center}
\end{table}

\end{document}